# Response of Carbon Nanotube (CNT) Ply Subjected to a Pulsed Magnetic Field

Ali Nassiri[1,2,*], Brad Kinsey[3]

1. Department of Materials Science and Engineering, The Ohio State University, 2041 College Road N, Columbus, OH 43210, USA
2. Center for Design and Manufacturing Excellence, The Ohio State University, 1314 Kinnear Road, Columbus, OH 43212, USA
3. Department of Mechanical Engineering, University of New Hampshire, 33 Academic Way, Durham, NH 03824, USA
* Corresponding author, Tel: +1(614)688-2594

Abstract: In this study, the possible deformation of a single Carbon Nanotube (CNT) ply subjected to a pulsed magnetic field was investigated. In all tests the capacitor bank was charged to 6kJ of energy. A Photon Doppler Velocimetry (PDV) system was used to measure velocity or displacement of the CNT ply during the experiments. The resistance of the CNT ply was measured using four-point probe technique before and after the experiments. Preliminary results show that the single CNT plies do not permanently deform in response to the pulsed magnetic fields. However, they can be displaced, either by themselves a small amount (0.6mm) or by a large amount using a driver material. Also, the resistance of the CNT plies may increase or decrease depends on the lay-out (i.e., yarn) and current flow directions.

## 1- Introduction

Carbon nanotubes is usually considered the ultimate carbon fibers. The mechanical properties of nanotubes (i.e., strength and stiffness) is expected to approach that of an ideal carbon fiber, which has the perfect orientation of defect-free graphene layers along the fiber lay-out [1]. Theoretical predictions on single-walled nanotubes suggest that the Young's modulus should be close to the in-plane elastic modulus of graphite ~1060 GPa [2,3] and the value of tensile strength would be 200 GPa.

The addition of small quantities of CNTs to polymer composites is known to cause a dramatic increase in the thermal conductivity of the polymer host [4]. The thermal conductivity will change from around 0.1 to 1 W/mK for neat polymers, to as much as 10 W/mK for single-walled CNT composites.

The properties of noncomposite CNT assemblies may be further enhanced by magnetic alignment [5,6]. However, the theoretical limit to such assemblies is the thermal crosslinking between connected nanotubes which diffuses the thermal wave vector $k$ [7]. Hence, the goal of the present work has been to (1) investigate the possible deformation of a single CNT ply subjected to a pulsed magnetic field (2) measure the resistance of CNT ply in response to magnetic field.

During the experiments, a pulsed power supply discharges into a uniform pressure coil which creates a magnetic field in close proximity to the CNT ply. Eddy current are induced in the CNT, and a repulsive Lorentz force is created from the magnetic field- eddy current interaction which causes the CNT ply to accelerate away from the coil and deform CNT ply.

Impulse forming and welding are a cutting edge technology in advanced manufacturing [8,9]. For example, in magnetic pulsed welding (MPW), which is a solid-state welding process, the metallurgical bond between two dissimilar metals can be achieved without melting and solidification. Also, there is no heat-affected zone generated after welding. Hence, welded materials improve or retain their original properties. Moreover, the energy consumption and cost of MPW is moderate when compared to fusion welding methods.

## 2- Single CNT ply experiments

Figures 1-4 show the experimental setup for single CNT ply subjected to a pulsed magnetic field. A 12.7cm x 6.7cm CNT ply is placed on the uniform pressure coil that designed and fabricated by Thibaudeau and Kinsey [10] (Fig. 1). Note the white patch in the center of the CNT ply is photo

reflective tape for the Photon Doppler Velocimetry (PDV) system measurement [11,12,13].

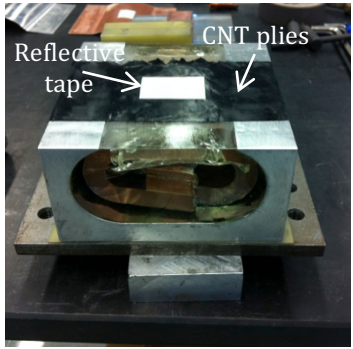

Fig. 1. CNT ply placed on the uniform pressure coil

The CNT ply is carefully stretched before the 25.4mm spacers with 10mm radius are used between the CNT ply and the clamp plate (Figs. 2-3). Figure 4 shows the experimental setup including the Aluminum forming box, Rogowski coil (for measuring current), PDV system, and oscilloscope.

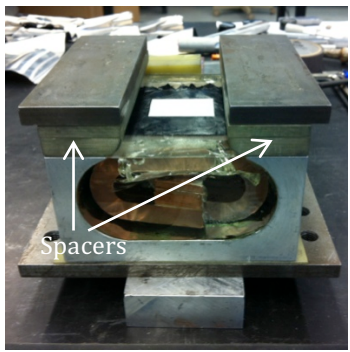

Fig. 2. Spacers on the CNT ply

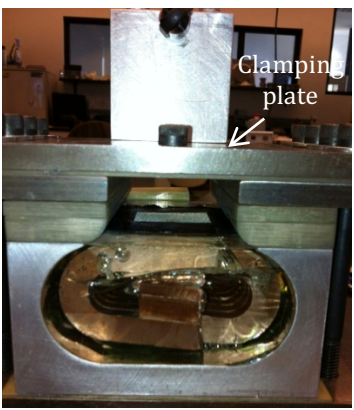

Fig. 3. The CNT ply is kept stretched by tightening bolts through the clamping plate

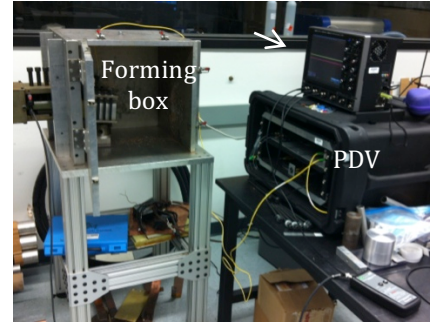

Fig. 4. Experimental setup

Fig. 5 shows the velocity of the CNT ply (the average of five tests) subjected to a pulsed magnetic field. The captured velocity is much less than equivalent experiments for the metal sheets as will be shown in a subsequent section. The velocity profile is also different with an oscillating behavior.

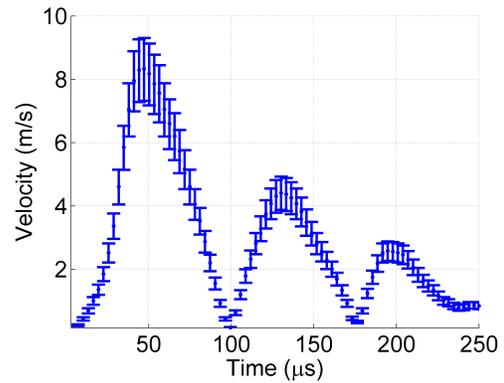

Fig.5. Velocity vs. time

Fig. 6 shows the displacement of CNT ply. The plot indicates that displacement is very small.

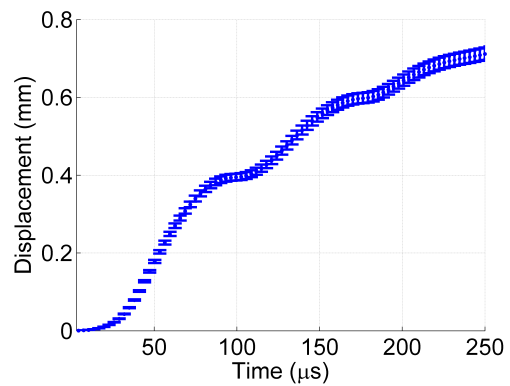

Fig.6. Displacement vs. time

Based on these tests, there are two observations of the final CNT ply after being subjected to the pulsed magnetic field (see Figs. 7, 8):

i- No sign of permanent deformation is observed for the single CNT ply.

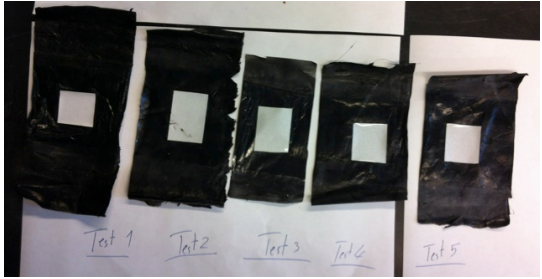

Fig. 7. CNT plies

ii- It seems the material properties are changed for the part of CNT ply which is subjected to the magnetic field.

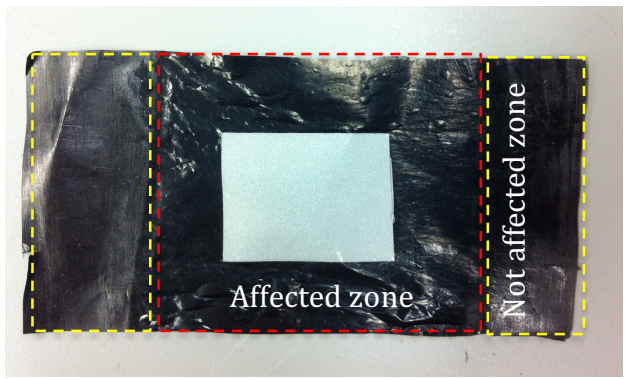

Fig. 8. Some effects on the surface of the region subjected to the magnetic field

To validate this hypothesis some basic tests, e.g. resistivity, are recommended which will be addressed in Section 4.

## 3- CNT ply with copper sheet driver

Similar experiments were conducted with a single CNT ply placed on top of a 1 mm Copper sheet (Cu 510) to use as a driver material. Figs. 9-12 show the experimental setup for this experiment.

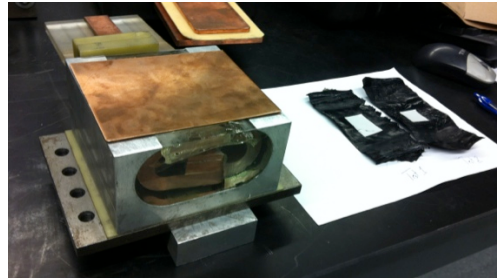

Fig. 9. Copper sheet on the uniform pressure coil

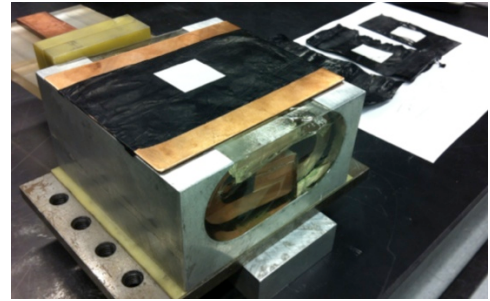

Fig. 10. CNT ply is placed on the Copper sheet

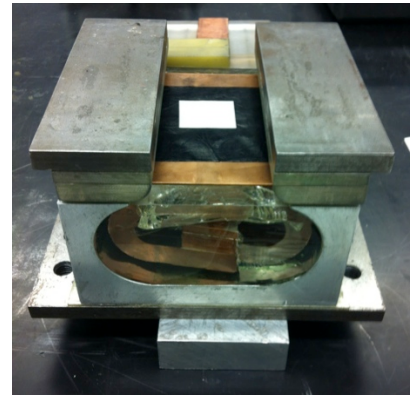

Fig. 11. Spacers on the CNT ply and Copper sheet

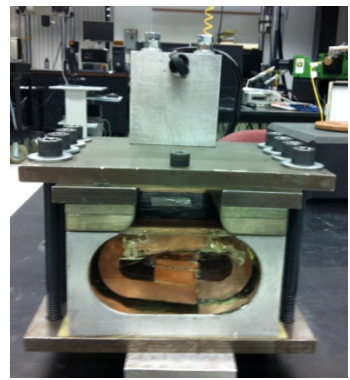

Fig. 12. Experimental set-up

Figs. 13, 14 show a forming result.

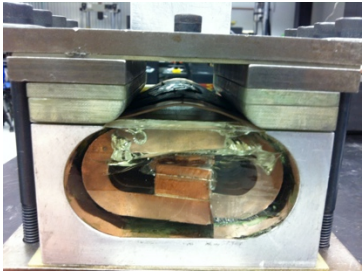

Fig. 13. Forming result

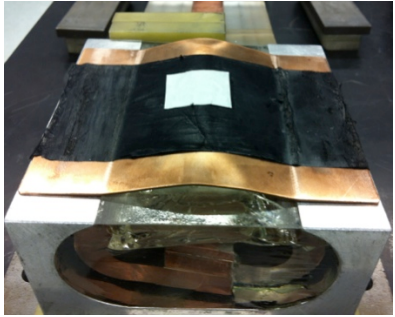

Fig. 14. Forming result

Fig. 15 shows the velocity of the CNT ply with driver subjected to the pulsed magnetic field. (Notice that the maximum velocity of the single CNT ply shown in Fig. 5 was 8 m/s, almost an order of magnitude less.)

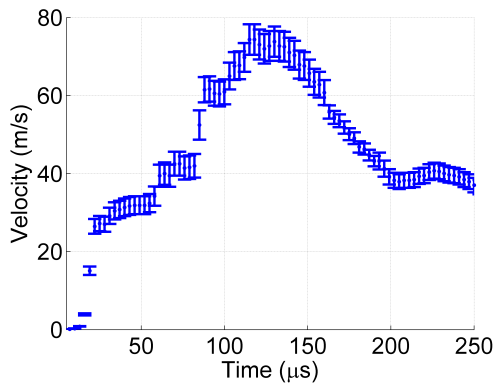

Fig. 15. Velocity vs. time

Fig. 16 shows the displacement of CNT ply with the driver material. (The single CNT ply only displaced by 0.6 mm, see Fig. 6.).

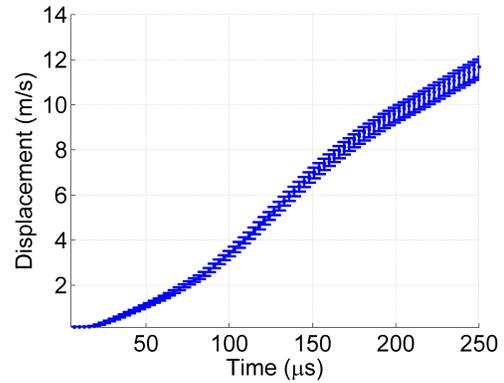

Fig. 16. Displacement vs. time

Based on these tests, there are two observations of the final CNT plies after being subjected to the pulsed magnetic field with the driver material:

i- No sign of permanent deformation is observed for the CNT ply.

ii- The captured data is consistent with our previous tests for the single Cu 510 sheet.

Fig.18 shows all the experimental test samples in this study.

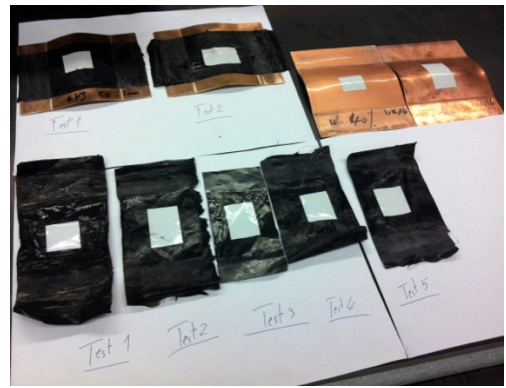

Fig. 18. All experimental tests

## 4- Resistance of Carbon Nanotube (CNT) plies subjected to pulsed magnetic fields

In this section, data from the measurement of resistance of Carbon Nanotube (CNT) plies (Lot Number: 70674, provided by Nanocomp) using the four-point probe technique is presented before and after a "single" and "three" magnetic pulses from a uniform pressure coil. In two of experiments, the CNT ply was six inches in length and four inches wide. In an additional experiment, the size was four inches in length and four inches wide to change the "lay-out direction" of the CNT ply with respect to the current flow. In all experiments the capacitor bank was charged to 10.8kJ of energy.

## 5- Experimental setup:

A 12-inch by 12-inch CNT ply was cut to 6-inch by 4-inch and 4-inch by 4-inch samples (See Fig. 19).

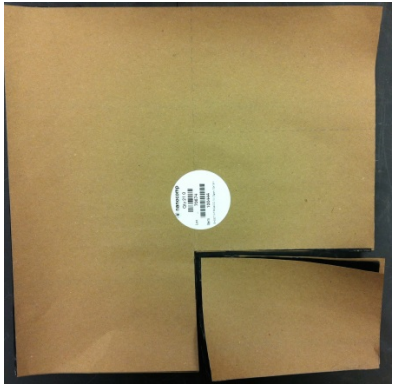

Fig. 19. CNT ply cut to 6" x 4" and 4"x4" based on the direction of barcode (i.e., CNT lay-out direction)

A schematic of the CNT ply and the current flow direction during tests are shown in Figs. 20 and 21. Note the direction of the CNT lay-out in Fig. 2. This was not specified by Nanocomp but inferred from resistance measurements. Also, note that the direction of the current flow indicated in Fig. 2 (e.g., left to right versus right to left) was not controlled during the tests.

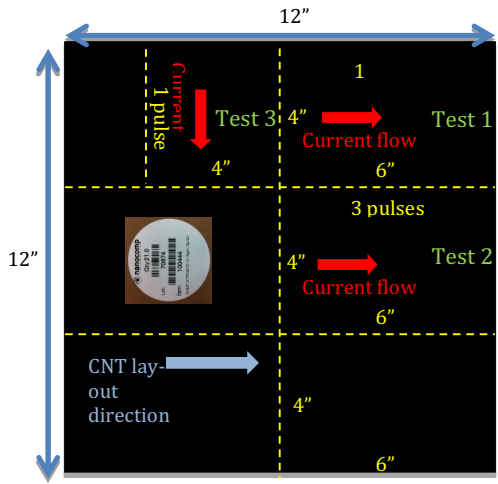

Fig. 20. Schematic of CNT ply and current flow direction during test

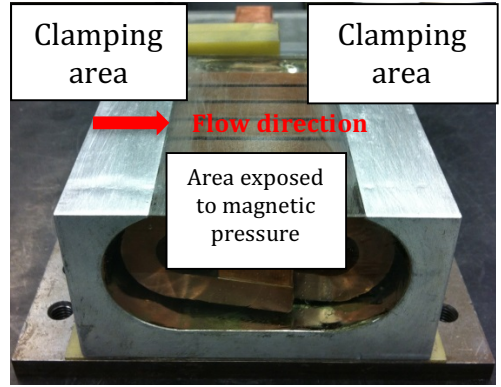

Fig. 21. Flow direction in uniform pressure coil

### Test 1:

In the first test, a 6" x 4" CNT ply was subjected to a "single" magnetic pulse. Fig. 22 shows the experimental setup and Fig. 23 shows the four point probe used to measure resistance with data presented in Figs. 24 and 25 before and after the test.

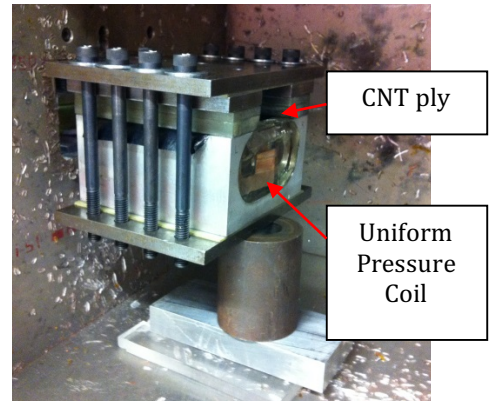

Fig. 22. Experimental set-up

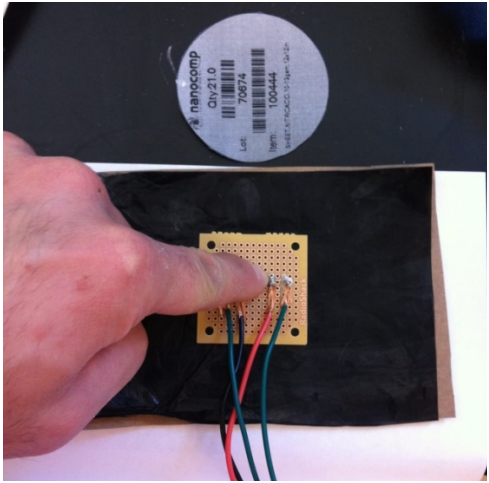

Fig. 23. Four-point probe resistance measurement along current flow direction and parallel to CNT lay-out direction for Tests 1 and 2

The slopes in Figs. 24 and 25 indicate the resistance values. For the same sample, the resistance was measured in the direction perpendicular to the CNT lay-out (i.e., perpendicular to the probe direction shown in Fig. 23) before and after a single magnetic pulse (data presented in Table 1). The resistance data for all tests are the average of ten sweeps, i.e., varying the current over a range and measuring the voltage.

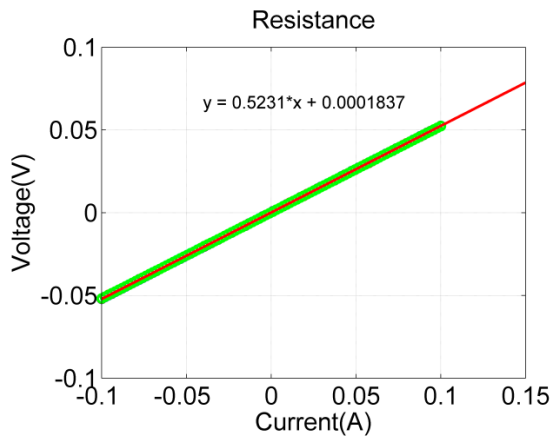

Fig. 24. CNT ply resistance "before" a single magnetic pulse with the probe direction as shown in Fig. 6

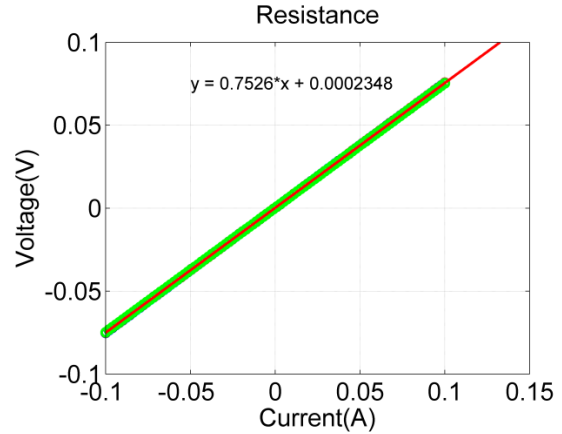

Fig. 25. CNT ply resistance "after" a single magnetic pulse (the data presented are the average of ten sweeps) with the probe direction as shown in Fig. 23

Results show that the resistance increased by 44% for a "single" magnetic pulse parallel to the CNT lay-out direction and increased by 54% perpendicular to the CNT lay-out direction.

Test 2:

In this second test, a CNT ply was subjected to "three" magnetic pulses (with the same setup as Test 1). See Table 1 for data before and after tests and Fig. 26 of picture while specimen was still in the experimental set-up.

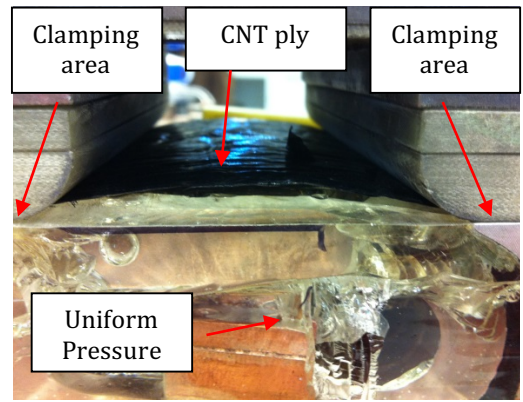

Fig. 26. CNT ply "after" three magnetic pulses

Results show that the resistance increased by 71% for "three" magnetic pulses parallel to the CNT lay-out direction and increased by 83% perpendicular to the CNT lay-out direction.

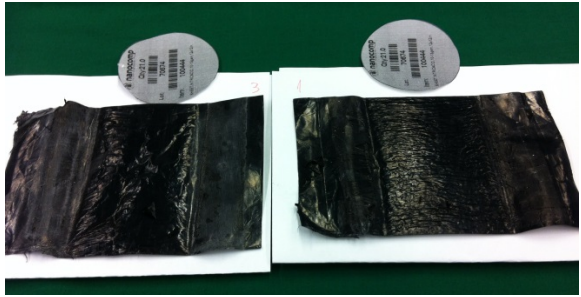

Fig. 27. CNT ply surfaces after three (left) and a single (right) magnetic pulse (Test 2 and Test 1 respectively)

Test 3:

In the third test, a 4-inch by 4-inch sample was subjected to a "single" magnetic pulse. In this test (unlike the first two tests) the current flow was perpendicular to CNT lay-out direction.

For this test, the resistance increased by 54% for a "single" magnetic pulse parallel to the CNT lay-out direction and decreased by 42% perpendicular to the CNT lay-out direction.

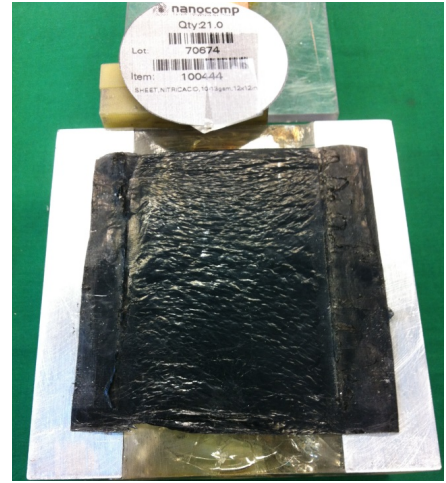

Fig. 28. 4"x4" CNT ply surfaces after a "single" magnetic pulse (Test 3)

Table 1. Resistance values for the various tests

|  | Number of pulses | Direction of current flow to CNT lay-out | Direction of resistance measurement to CNT lay-out | Before | After | % of change |
|---|---|---|---|---|---|---|
| **Test 1** | 1 | Parallel | Parallel | 0.5231 | 0.7526 | 44% |
|  |  |  | Perpendicular | 0.6939 | 1.073 | 54% |
| **Test 2** | 3 | Parallel | Parallel | 0.4683 | 0.8016 | 71% |
|  |  |  | Perpendicular | 0.5906 | 1.081 | 83% |
| **Test 3** | 1 | Perpendicular | Parallel | 0.146 | 0.2246 | 54% |
|  |  |  | Perpendicular | 0.3684 | 0.2124 | -42% |

## 6- Preliminary Results:

- The single CNT plies do not permanently deform in response to the pulsed magnetic fields. However, they can be displaced, either by themselves a small amount (0.6 mm) or by a larger amount using a driver material.
- The material properties are changed by exposure to the pulsed magnetic field so it is possible that this method could be used to tailor properties of the final material.
- A recommendation if permanent deformation is of interest is to create a composite material with multiple CNT plies that are subjected to pulsed magnetic field.
- The data for all of these tests, both parallel and perpendicular to the CNT lay-out directions, are presented in Table 1. As this data indicates, the resistance increased in both directions when the current flow is parallel to the CNT lay-out direction after the magnetic pulses are applied. This can be explained by the deformation, albeit slight, that occurs in the material due to the magnetic pulse. Note that the increase was higher for additional pulses. Alternatively, when the specimen was oriented such that the CNT lay-out was perpendicular to the current flow direction, the resistance decreased instead of increased in the perpendicular direction to the CNT lay-out. This could be due to the electrical current flowing in this direction and thus better aligning the CNTs. Thus, the anisotropic resistivity/conductivity of the materials can be manipulated accordingly.
- The resistance values (as opposed to the resistivity values) are presented in this study. Since there is a direct correlation between resistance and resistivity (related by the length, width and thickness of the sheet), the resistivity is affected with these same rates even though only resistance change was documented.
- Note that since resistance is changed, the conductivity of CNT ply will be affected as well at similar rates if temperature effects are neglected.
- Additional tests where three pulses were applied with the current flow perpendicular to the CNT lay-out direction were attempted. However, failure of the epoxy in the Uniform Pressure Coil (see Fig. 26 where epoxy cracking is evident) prevented these tests from being completed. The Uniform Pressure Coil will be repotted with epoxy shortly so tests can continue. However, results from Tests 1 and 2 indicate that further changes in resistance values are possible with additional magnetic pulses.
- More statistical analyses are required for these tests as the resistance values are affected by the location of the samples on the full sheet and measurements.
- In all tests, the probes were placed at the center of CNT ply to measure resistance.